# Investigating the agreement between methods of different precision

Magnus Borga, PhD

Linköping university, Sweden

**Abstract**

Agreement between measurement methods is commonly investigated by a so-called Bland-Altman plot showing if the difference is independent of the size of the measurement. However, such analysis assumes that both methods have the same precision. If not, the plot may show a false trend caused by the difference in precision and not by an actual dependence on the size of the measurement. We suggest a modification of the Bland-Altman plot such that the differences are plotted against an inverse-variance weighted average of the measurements rather than their mean. This study shows that such modification removes the dependence on difference in precision.

# Background

In method comparison studies, it is common to investigate the agreement by a so-called Bland-Altman plot[1,2]. The Bland-Altman plot shows the differences between the methods' measurements plotted against the mean of their measurements. From this analysis, the bias and limits of agreement between the methods can be determined. Of particular interest is to investigate if there is a linear trend in the bias, in other words if the difference between methods depends on the magnitude of the measurements. This is important when generalizing the results of a method comparison study to other cohorts with a potentially different distribution of the measurement values. If, for example, the agreement between two methods for measuring blood pressure is evaluated on a randomly selected cohort from the general population and this evaluation shows a linear trend in the bias, the estimated bias and limits of agreement will not be valid on a cohort with high blood pressure unless this linear trend is accounted for.

A limitation with this analysis method is that it is only valid when the measurement errors of the two methods have approximately the same variance, i.e. that both methods have the same precision. If the methods have different precision, this will cause a linear trend in the Bland-Altman plot even if there is no linear trend in the measurement errors. Therefore, an observed linear trend in the Bland-Altman plot can be a consequence of either a true linear dependence between the differences and the magnitude of the measurements, or a difference in precision between the methods (or a combination of both). Furthermore, the absence of a linear trend does not guarantee that the difference between measurements is independent of the magnitude of the measurements since the two effects may cancel each other!

An intuitive way of understanding this effect is to consider the effect of the measurement noise on a particular measurement point when *a - b* is plotted against (*a+b*)/2, *a* and *b* being measurements from two different methods or devises. The distribution of a measurement error in *a* will be along a line with a positive slope while the corresponding distribution for a measurement error in *b* will have a negative slope. Only if these error distributions are equal, these effects will cancel each other and the combined effect of the two measurement errors will be unbiased.

This limitation was recognized already by Altman and Bland[3] and then re-discovered and further investigated by others.[4] However, to the best of our knowledge, no solution has been presented to enable analysis of the trend in the bias for methods with different precision.

To overcome this problem, we suggest a modification of the Bland-Altman analysis which generalizes its application also to comparisons of methods with different precision. The modification is to replace the arithmetic mean plotted on the horizontal axis by a weighted average where the weights are given by the reciprocal of the within-subject variance for each method respectively. Or, equivalently, each method's measurements should be weighted by the other method's within-subject variance:

$$\frac{s_{wB}^2 a + s_{wA}^2 b}{s_{wA}^2 + s_{wB}^2}$$

where *a* and *b* are the measurements and $s_{wA}^2$ and $s_{wB}^2$ are the within-subject sample variances for methods A and B respectively. Note that, in the case where $s_{wA}^2 = s_{wB}^2$, this is equivalent to the arithmetic mean, $(a + b)/2$. Hence, this generalization is equivalent to the standard formulation of the Bland-Altman analysis when the condition for such analysis is fulfilled. But, in contrast to the standard Bland-Altman analysis, this analysis method is valid also when the precision of the methods differ.

In the following sections we show analytically that the proposed modification indeed removes the linear bias component caused by differences in within-subject variance between the methods. We also illustrate the effects of true linear dependency and differences in precision on Bland-Altman plots of synthetic data with and without the proposed modification. Finally, we show the effect of this modification on real data from an article by Bland and Altman.[5]

## Methods

We first show analytically that the proposed method removes any correlation between the difference and weighted average when the true difference is uncorrelated to the magnitude of the measurement.

In order to illustrate the effects of differences in method precision in the standard Bland-Altman analysis and how this these effects are avoided by the proposed method, we generated synthetic data where the measurement bias and precision could be controlled. Three uncorrelated signals with 100 sample points each were generated from a normal distribution. One was used as a common underlying signal $c$ (which could be viewed as an unknown ground truth), and the other two were used as independent measurement errors $\varepsilon_A$ and $\varepsilon_B$. The measurements $a$ and $b$ were then calculated as

$$a = k_A \cdot c + s_A \cdot \varepsilon_A \text{ and}$$

$$b = k_B \cdot c + s_B \cdot \varepsilon_B$$

where $k_A$ and $k_B$ are linear scaling factors of the common signal, and $s_A$ and $s_B$ are the sample standard deviations of the measurement errors. To ensure that linear independence of $c, \varepsilon_A$ and $\varepsilon_B$, as well as unit sample variance were reflected in the sample, the generated data were de-correlated and normalized using principal component analysis. Four different cases were generated, two with the same sample standard deviations of the measurement errors ($s_A = s_B = 1.5$) and two with different sample standard deviations ($s_A = 0.5$, $s_B = 4.5$). In each of these pair of cases, one case had equal scaling factors ($k_A = k_B = 1.0$) and the other one had different scaling factors ($k_A = 1.0$, $k_B = 0.9$). All these four cases were analysed using the standard Bland-Altman plot and the proposed method. The Pearson correlation coefficient, p-value for correlation, and estimated linear component in the difference with 95% confidence interval were then computed for the different cases.

Finally, to illustrate the effect of the proposed method on real data we have applied it on a data set used by Bland and Altman.[5] The data are adopted from Table 1 in the same article and shows blood pressure measurements made by an experienced observer (J) using a sphygmomanometer and by a semi-automatic blood pressure monitor (S). Each measurement was repeated three times which allowed for the within-subject variance to be determined. The correlation and the slope of a linear fit to the difference were calculated both for the standard Bland-Altman plot and for the proposed method.

All simulations and analyses were performed in Matlab R 2019a (The MathWorks, Inc., Natick, Massachusetts, United States.)

# Results

## Analytical results

Suppose there is no underlying dependence between the differences and the magnitude of the measurement methods A and B, here represented by the stochastic variables *A* and *B*. In that case, we want the correlation between the difference and the weighted average to be zero or, equivalently, the covariance to be zero. It can be shown (see Appendix 1) that, under this condition, the covariance between the difference and a weighted average is

$$\text{cov}\left(A - B, \frac{\alpha A + \beta B}{\alpha + \beta}\right) = \frac{\alpha \sigma_{wA}^2 - \beta \sigma_{wB}^2}{\alpha + \beta}, \qquad (1)$$

where $\sigma_{wA}^2$ and $\sigma_{wB}^2$ are the within-subject variance of *A* and *B* respectively, and $\alpha$ and $\beta$ are the weights in a weighted average of the two variables. By setting $\alpha = \sigma_{wB}^2$ and $\beta = \sigma_{wA}^2$, the covariance (and also the correlation) will be zero.

We can also see from Equation 1, by setting $\alpha = \beta = 1$, that the standard Bland-Altman plot indeed gives a covariance, i.e. linear dependency, between the average and the difference if the within-subject variances of the two methods are not equal.

## Results on synthetic data

Figure 1 shows the standard Bland-Altman plots of the synthetic data and Figure 2 shows the corresponding plots using the proposed method on the same data. In both figures, panels (a) and (b) (upper row) show the cases where $s_A$ and $s_B$ are equal, i.e., where the condition for the standard Bland-Altman plot is fulfilled. Plots (c) and (d) (lower row) show the cases where the standard deviations differ, i.e. where the condition for the standard Bland-Altman plot is broken. In plots (a) and (c) (left column), there is no linear component in any of the measurement errors, which means that the bias should be constant. In plots (b) and (d) (right column), there is a negative trend in method B which should result in a negative trend in the bias. We see in Figure 2 that the slope of the bias is correctly estimated by the proposed method, even in panels (c) and (d) where the methods have different precision, which is not the case for the standard Bland-Altman plots in Figure 1. The correlations, p-values, and estimated linear slopes with 95% confidence intervals are summarized in Table 1.

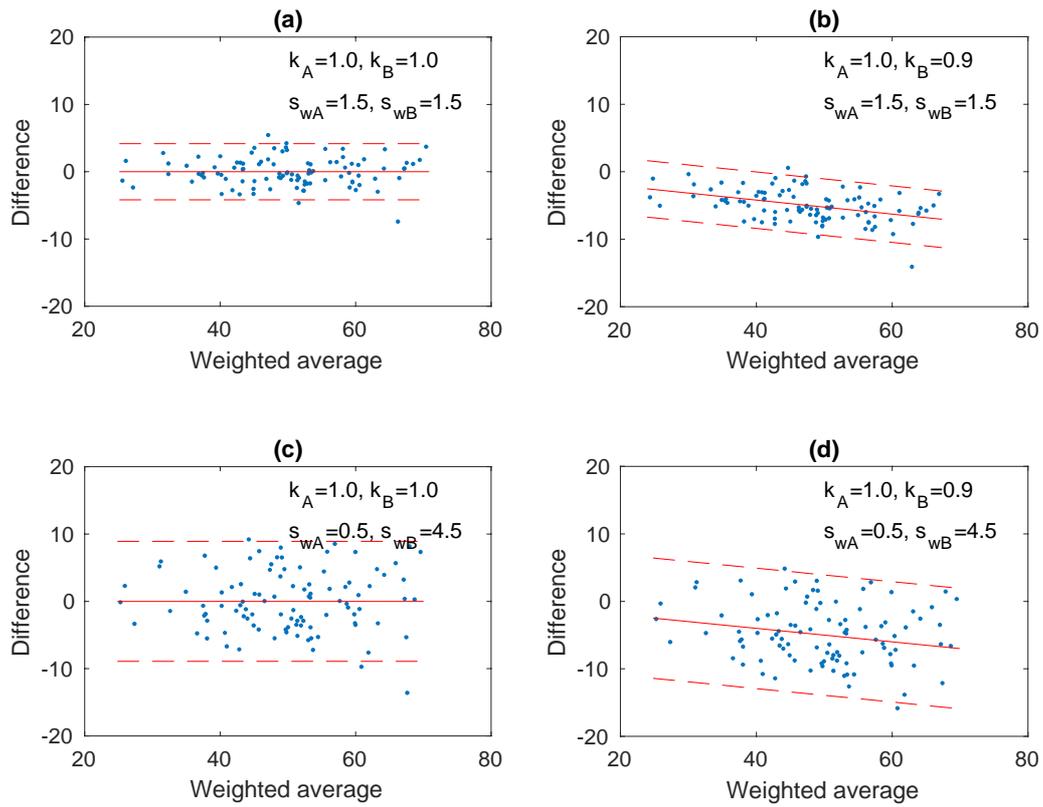

*Figure 1. Bland-Altman plots of synthetic data. Solid line indicates the bias and dashed lines indicate limits of agreement. Upper row: Both methods have the same precision. Lower row: Methods have different precision. Left column: No linear dependence in any of the measurement errors. Right column: Linearly dependent measurement error in one method.*

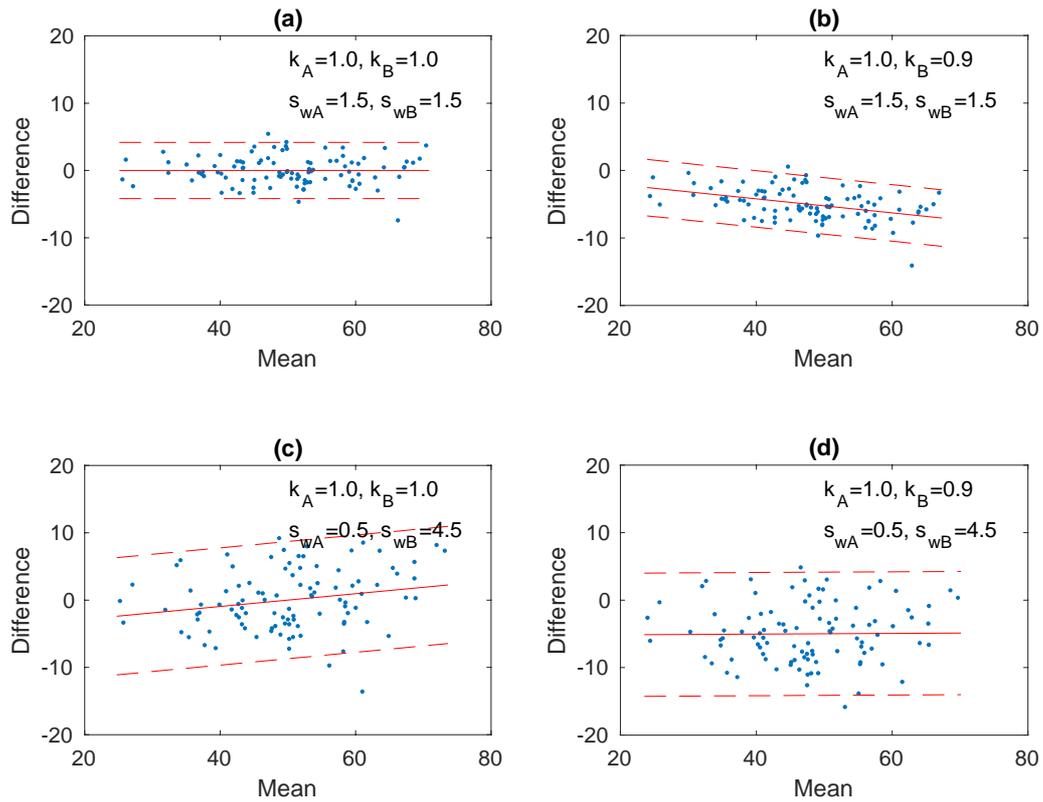

*Figure 2. Plots of the same synthetic data as in Figure 1 but using the proposed generalised method. Upper row: Both methods have the same precision. Lower row: Methods have different precision. Left column: No linear dependence in any of the measurement errors. Right column: Linearly dependent measurement error in one method.*

|  | Bland-Altman (Fig. 1) | | | Proposed method (Fig. 2) | | |
| --- | --- | --- | --- | --- | --- | --- |
| Case | $r$ | $p$ | $k$ | $r$ | $p$ | $k$ |
| a | 0.00 | 1.00 | 0.0 (-0.04 – 0.04) | 0.00 | 1.00 | 0.00 (-0.04 – 0.04) |
| b | -0.42 | <0.001 | -0.10 (-0.15 – -0.06) | -0.42 | <0.001 | -0.10 (-0.15 – 0.06) |
| c | 0.22 | 0.03 | 0.10 (0.01 – 0.18) | 0.00 | 1.00 | 0.00 (-0.09 – 0.09) |
| d | 0.01 | 0.91 | 0.01 (-0.09 – 0.10) | -0.22 | 0.03 | -0.10 (-0.19 – -0.01) |

*Table 1. Pearson correlation coefficient (r), p-value for correlation (p), and estimated linear component in the difference (k) for the different cases in figures 1 and 2.*

## Results on real data

The standard Bland-Altman plot of the blood pressure data is shown to the left in Figure 3 and to the right is the corresponding plot using the proposed method where the weighted average on the horizontal axis is calculated as

$$\frac{s_{wJ}^2 \cdot S + s_{wS}^2 \cdot J}{s_{wJ}^2 + s_{wS}^2},$$

where $s_{wJ}^2 = 37.4$ and $s_{wS}^2 = 83.1$. The estimated slope of the bias was k=0.01 (-0.13 – 0.14) with the proposed method and -0.07 (-0.21 – 0.07) with the Bland-Altman plot. In this example, none of the methods can reject the hypothesis that there is no linear dependence between the differences and the blood pressure, but the trend almost completely disappears with the proposed method.

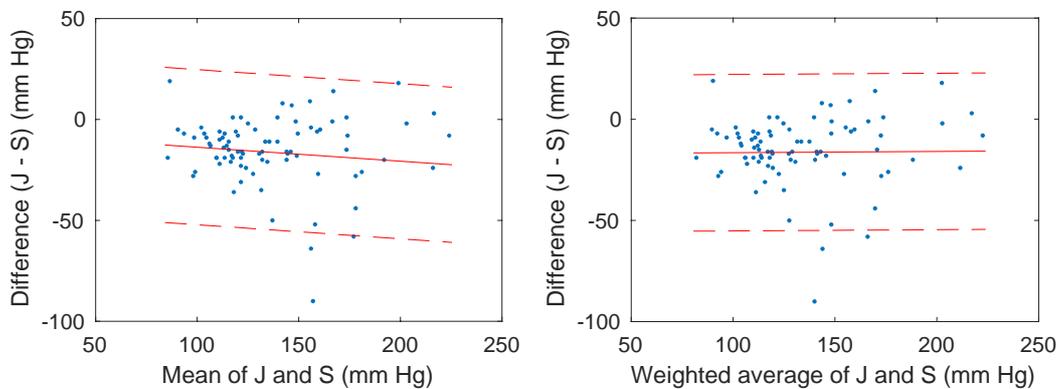

*Figure 3. Standard Bland-Altman plot (left) and the corresponding plot using the proposed method (right) for blood pressure data adopted from Bland and Altman 1999.[5] The negative linear component apparent in the original plot has almost vanished in the proposed plot.*

## Conclusions

We have presented a method for investigating the agreement between different measurement methods that can be seen as a generalization of the standard Bland-Altman plot. When the assumptions for the Bland-Altman plot are fulfilled, i.e. when the methods' precisions are equal, the plots become identical. But while a difference in method precision introduces a linear trend in the bias of the Bland-Altman plot, the proposed method is unaffected by differences in method precision and any slope of the bias is solely caused by a dependence between the difference between measurements and the magnitude of the measurements.

One way of understanding this method is to view the weighted average as an optimal (in a least mean square error sense) estimate of the ground truth, obtained by weighting the measurements with the reciprocals of their variances, in statistics known as "inverse-variance weighting".[6] A consequence of this is that when the variance of the measurement error of one of the methods approaches zero, the weighted average approaches that method's measurements. In other words, if one of the methods would be identical to the ground truth without any measurement error, the proposed method would plot the difference against the ground truth. Assume method B is such an ideal method without any measurement error. The covariance between $a - b$ and $b$ can be found by setting $\alpha = 0$ in Equation 1, which gives

$\text{cov}(a - b, b) = -\sigma_{wB}^2$ (if there is no underlying dependence between the differences and the magnitude of the measurements). But as $\sigma_{wB}^2$ approaches zero, so does the covariance, and therefore also the correlation between $a - b$ and $b$. At a first glance, this might seem to contradict the conclusions by Bland and Altman[3] where they showed that plotting the difference against the standard method indeed introduces a slope in the bias. However, this is no contradiction since the "standard method" is not the same as the ground truth. In practice, of course, this will never occur. Any physical measurement has a certain measurement error and the underlying ground truth is always unknown. However, in simulations using synthetic data, where the ground truth is actually known, this shows that in such case, the difference should actually be plotted against the ground truth. This also explains the conclusion by Krouwer that one "should use *X*, not (*Y*+*X*)/2 when *X* is a reference method".[7] In that case, the so-called reference method has a much higher precision and therefore a negligible error variance compared to the other method, so an inverse-variance weighted average would be almost equal to the reference method.

A limitation with the proposed method is that it requires the determination of the precision of both methods. While this is always desirable from a statistical perspective, it can sometimes be unfeasible due to practical or ethical reasons to obtain multiple measurements on patients. Without multiple measurements, we are in general not able to determine if a linear trend in the bias is caused by a linear dependency between the errors and the true value, or if it is caused by a difference in within-subject variance between the methods.[8] However, if the precisions of both methods are known from other sources, we suggest that this information is used to produce the weighted average, rather than assuming equal precession. Another limitation is that this framework does not cover the case where the precision in any of the methods is dependent on the size of the measurements.

In conclusion we suggest that, in method comparison studies, one should always strive to determine the within-subject variance for each method, and use these estimated variances in a weighted average, replacing the mean on the horizontal axis in a Bland-Altman plot.

## Conflict of interest

The Author declares that there is no conflict of interest

## Funding

This work was funded by the Swedish Research Council [VR 2019-0475].

# Appendix 1

## Proof of Equation 1

The covariance between the difference and the weighted average of the stochastic variables $A$ and $B$ is

$$\text{cov}\left(A - B, \frac{\alpha A + \beta B}{\alpha + \beta}\right) = \frac{1}{\alpha + \beta}\left(\alpha \sigma_A^2 - \beta \sigma_B^2 + (\beta - \alpha)\text{cov}(A, B)\right),$$

where $\sigma_A^2$ and $\sigma_B^2$ are the variances of the distributions of $A$ and $B$ respectively. Now let us divide each variable into two components, one proportional to an underlying common stochastic variable $C$, and another which is a measurement error $\varepsilon$, uncorrelated to $C$:

$$A = k_A \cdot C + \varepsilon_A \text{ and } B = k_B \cdot C + \varepsilon_B,$$

where $k_A$ and $k_B$ are scaling factors for each variable. Since we assumed that there is no underlying dependence between the differences and the magnitude of the measurements, the scaling factors $k_A$ and $k_B$ must be equal, i.e., $k_A = k_B = k$, which means that $A$ and $B$ can be written as

$$A = k \cdot C + \varepsilon_A \text{ and } B = k \cdot C + \varepsilon_B.$$

The variance of each measurement method can then be written as

$$\sigma_A^2 = k^2 \cdot \sigma_C^2 + \sigma_{wA}^2 \text{ and } \sigma_A^2 = k^2 \cdot \sigma_C^2 + \sigma_{wB}^2,$$

where $\sigma_C^2$ is the variance of the distribution of $C$ and $\sigma_{wA}^2$ and $\sigma_{wB}^2$ are the variances of the independent measurement errors, i.e., the within-subject variance. Since $\varepsilon_A$ and $\varepsilon_B$ are independent, this means that

$$\text{cov}(A, B) = k^2 \sigma_C^2.$$

Hence,

$$\text{cov}\left(A - B, \frac{\alpha A + \beta B}{\alpha + \beta}\right) = \frac{1}{\alpha + \beta}\left(\alpha(k^2\sigma_C^2 + \sigma_{wA}^2) - \beta(k^2\sigma_C^2 + \sigma_{wB}^2) + (\beta - \alpha)k^2\sigma_C^2\right)$$
$$= \frac{\alpha \sigma_{wA}^2 - \beta \sigma_{wB}^2}{\alpha + \beta} \quad \therefore$$